\documentstyle[editedvolume,psfig]{crckapb10} 

\begin{opening}
\title{A complete sample of quasars from the 7C redshift survey}

\author{Chris J. Willott}
\author{Steve Rawlings}
\author{Katherine M. Blundell}
\author{Mark Lacy}
\institute{Astrophysics,Department of Physics,Keble Road,Oxford OX1 3RH}

\end{opening}

\runningtitle{7C quasar sample}

\begin{document}

\begin{abstract}{\small
 
We present details of a new sample of radio-loud quasars drawn from 0.013 sr of the 7C Redshift Survey. This sample is small (21 quasars) but {\em complete} in that every object with an unresolved nucleus and/or broad emission lines with $S_{151MHz} > 0.5$ Jy has been discovered. The dependence of the quasar fraction with redshift and radio luminosity is investigated, providing new evidence supporting the unification of radio-loud quasars and powerful radio galaxies. This 7C sample is compared with optically-selected quasars, in order to determine whether there are systematic biases in the different selection techniques. There are no lightly reddened ($A{_{V}} \sim 1$) quasars in our sample amongst the 14 with $z < 2$. The discovery of a reddened quasar at $z = 2.034$ and its implications are discussed. A tight correlation between radio luminosity and optical/near infrared continuum luminosity for a subset of the sample is also found.
}
\end{abstract} 

\section{Introduction}

Quasars are the most distant objects known, so their study is a fundamental tool in cosmology. Their strong, broad ($FWHM > 2000$ km s$^{-1}$) emission lines and blue continua make their identification relatively easy. Traditionally, quasar searches were made in the optical by identifying unresolved blue objects on photographic plates. The problem with such searches is that both a limiting plate magnitude and colour selection exclude intrinsically faint (and/or reddened) quasars. However, approximately one-in-ten quasars are also strong radio sources and this property can be used to select quasar samples. Even in these radio-selected samples, an optical selection criterion, such as a limiting magnitude, is frequently imposed. The sample of 7C quasars described here attempts to avoid such biases in an effort to understand quasars, and particularly, their relationship with powerful radio galaxies.       

\section{The 7C Redshift Survey}

For several years, we have been pursuing a programme to obtain redshifts for every member of a sample of 7C radio sources (see Rawlings et al., these proceedings) with 151 MHz flux density $S_{151} > 0.5$ Jy in three regions of the sky. This work is now nearing completion and full results are to be published soon. VLA (predominantly A-array) maps (Blundell et al. 1997) and K-band imaging (Willott et al. 1997a) have been acquired for nearly all of the 79 sources in two of these regions (totalling 0.013 sr). Accurate radio and near-IR positions of the sources have been used for extensive spectroscopic follow-up in both the optical and infrared (Willott et al. 1997a). 78 of the 79 objects have been identified with an optical/infrared counterpart. Secure spectroscopic redshifts have been obtained for 61 of these sources, giving a current redshift completion of 78\%. This is expected to rise to 90\% (the value in the most studied of the two regions) with scheduled spectroscopy of the second region.

The selection at the {\em low} radio frequency of 151 MHz of our sample is crucial because it minimises the relative strength of the flat-spectrum radio core to the extended steep-spectrum emission. This reduces the biases found in high-frequency selected samples due to beamed, Doppler-boosted, radio cores. The faint flux limit of this sample means that it can include moderately powerful radio sources out to high redshifts (see Section 3). Perhaps the most important property of the 7C Redshift Survey is that it is a complete sample of identified radio sources. There are no optical selection criteria. This is essential if comparisons of the properties of radio galaxies and quasars are to be statistically meaningful. Of the 79 sources in the sample, 21 are identified as quasars with $0.9 < z < 3.0$. An object is classified as a quasar if it has an unresolved, `stellar' optical ID and/or broad ($FWHM > 2000$ km s$^{-1}$) emission lines in its optical spectrum. We believe this to be a statistically complete sample of radio-loud quasars, because all the sources currently lacking spectroscopic redshifts have resolved optical/infrared IDs and are therefore radio galaxies.             

\section{The 7C Quasar Fraction}

According to the orientation-based AGN unification scheme proposed by Scheuer (1987) and Barthel (1989), quasars and powerful radio galaxies are the same objects viewed at different orientations. Radio galaxies are simply quasars viewed close to the plane of the sky, so that the central emission region is obscured by a dusty torus perpendicular to the radio axis (see Antonucci, 1993). The fraction of quasars in complete samples and their linear size distributions are tests of this hypothesis. For details of the linear size evolution of this sample see Blundell et al. (these proceedings). Here we will discuss the quasar fraction in the 7C sample and compare with results from the revised 3CR sample (Laing et al. 1983).

\begin{figure}
\vspace{-1.7cm}
\centerline{
\psfig{figure=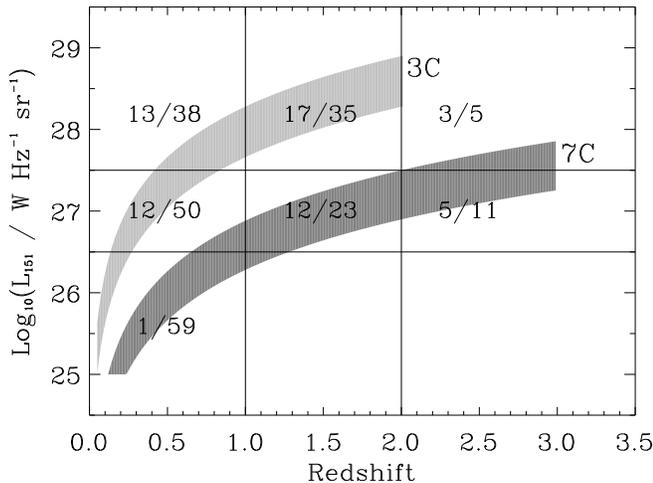,width=12.5cm}
}
\vspace{-0.9cm}
\caption{\label{fig} 
The radio luminosity - redshift plane. The regions containing the majority of the sources in the 3CR and 7C samples are shaded. The fraction of sources which are quasars are shown in each bin.} 
\vspace{-0.3cm}
\end{figure}

Figure 1. shows the regions of the luminosity-redshift plane occupied by the 3CR and 7C samples. The plane has been divided crudely into 9 bins to examine how the quasar fraction depends upon redshift and low-frequency radio luminosity. The first thing to notice is that there are virtually no quasars with radio luminosities below $log_{10}(L_{151}) \sim 26.5$. This has been noticed before (e.g. Kellermann et al. 1989) and implies a breakdown in simple unification-by-orientation at low luminosities. For the intermediate luminosity sources, the quasar fraction doubles from 24\% in the $0 < z < 1.0$ bin (predominantly 3CR) to 52\% in the $1.0 < z < 2.0$ bin (7C). This may be showing some differential evolution between the two populations, or may mean that some fraction of the radio galaxy population (e.g. those with low-excitation optical emission line spectra) does not unify. The space density of the radio source population increases by more than an order of magnitude over this redshift range (Dunlop \& Peacock 1990), so this relatively small change in the quasar fraction is certainly in accord with unified schemes. Now we consider the luminosity dependence of the quasar fraction in the range $1.0 < z < 2.0$. The quasar fraction of powerful sources ($log_{10}(L_{151}) > 27.5$) is the same as that at lower radio powers ($26.5 < log_{10}(L_{151}) < 27.5$). Hence there is no evidence that the quasar fraction depends upon the 151 MHz luminosity (provided $log_{10}(L_{151}) > 26.5$). For the simple model of Barthel (1989), this fraction of 50\% implies a cone opening angle to the line-of-sight of 60$^{\circ}$. Similar quasar fractions of $\sim$ 50\% are seen at higher redshifts ($z  > 2.0$), but the small number of objects in our sample at these redshifts prevents firm conclusions from being drawn.

\section{Comparison with Optically-Selected Quasars}

\begin{figure}
\vspace{-0.2cm}
\centerline{
\psfig{figure=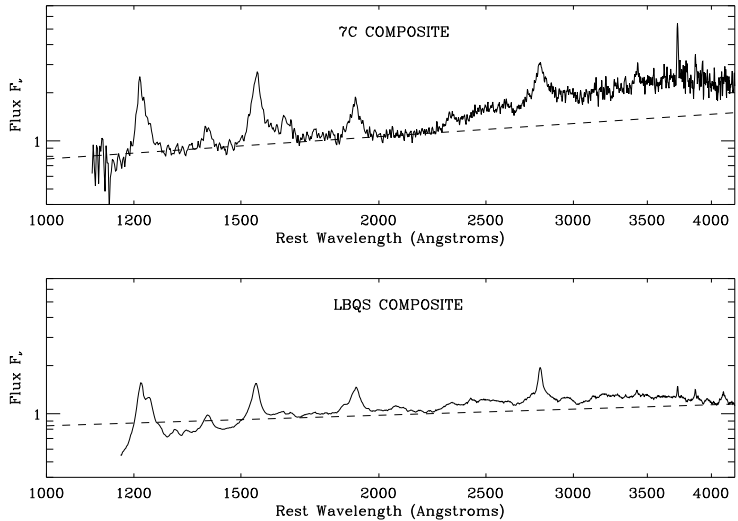,width=10.5cm}
}
\caption{\label{fig} 
Composite quasar spectra from the 7C sample (top) and the LBQS sample of Francis et al. (1992) (bottom). The dashed lines show power-law fits to the continuum with $\alpha{_{\rm opt}} = 0.5$ (7C) and $\alpha{_{\rm opt}} = 0.2$ (LBQS). Note that for $\lambda{_{\rm rest}} > 2300$ \AA\, the 3000 \AA\ bump adds another component to the continuum, which is more prominent in the 7C quasars.} 
\vspace{-0.2cm}
\end{figure}

Spectra of 16 7C quasars have been combined to create a composite quasar spectrum. Objects with poor signal-to-noise and the reddened cases discussed in Section 6 were omitted. This composite quasar spectrum was compared to the optically-selected LBQS composite of Francis et al. (1992). Figure 2 shows the 7C and LBQS composite quasar spectra. Note that the two spectra are qualitatively very similar, both in line equivalent widths and in spectral shape. The optical spectral index ($\alpha{_{\rm opt}}$, where $F{_\nu} \propto \nu^{-\alpha{_{\rm opt}}}$) of the 7C composite quasar is 0.5, while for the LBQS composite $\alpha{_{\rm opt}} = 0.2$. This may indicate that there is slightly more reddening in the 7C quasars than the optically-selected quasars, but we do not see any lightly reddened ($A{_{V}} \sim 1$) quasars, barring those discussed in Section 6. The population of lightly reddened quasars in 3CR is also small (see Rawlings et al., these proceedings).   

\section{Radio - Optical Correlations}

\begin{figure}
\vspace{-0.6cm}
\centerline{
\psfig{figure=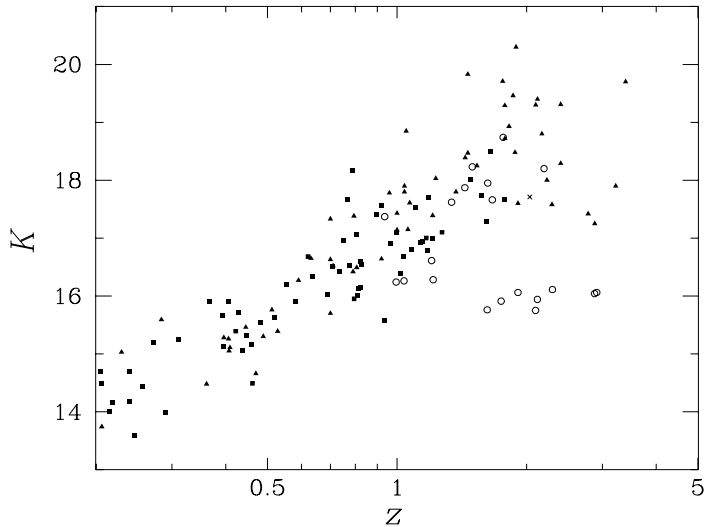,width=10.5cm,angle=270.0}
}
\vspace{-0.3cm}
\caption{\label{fig} 
Near infrared $2.2 \mu m\ K$-magnitude against redshift, $z$. 3C and 6C radio galaxies are plotted as filled squares and triangles, respectively. The open circles are the 7C quasar sample described here. The reddened quasar, 5C7.195 (see Section 6), is shown as a cross.  } 
\vspace{-0.3cm}
\end{figure}

Figure 3 plots the near-infrared $K$ magnitudes of the 7C quasars, along with 3C and 6C radio galaxies (Eales \& Rawlings 1996), as a function of $z$. It is immediately apparent that the quasars form two distinct groups in this diagram. 9 of the quasars lie amongst the radio galaxies along the $K-z$ relation (Lilly \& Longair 1984). The other 12 7C quasars all have similar $K$ magnitudes ($K = 16.1 \pm 0.5$), despite redshifts ranging from 1 to 3. This lower group therefore shows a tight correlation between the rest-frame optical/NIR continuum luminosity and redshift, and/or (since this is a flux-limited sample) radio luminosity. Intuitively, the correlation with radio luminosity seems more likely because the radio and optical luminosities both draw their energy from the same central engine. Indirect evidence for such a relation (Rawlings \& Saunders 1991) has recently been supported by studies of a larger quasar sample by Serjeant et al. (1997). The separation of these two groups in the $K-z$ diagram is not due to reddening effects because the same `bimodal' distribution is seen in the $B{_{J}}-z$ plot and we have already seen (Section 4) that the quasar spectra are not significantly reddened. For the quasars which have $K$ magnitudes similar to radio galaxies at the same redshift, the central quasar emission is relatively weak in K-band. However, all of these quasars are unresolved in K-band, so the quasar light still dominates over starlight.   

\section{A Reddened Quasar at $z \sim 2$}

\begin{figure}[tbh]
\vspace{-0.5cm}
\centerline{
\psfig{figure=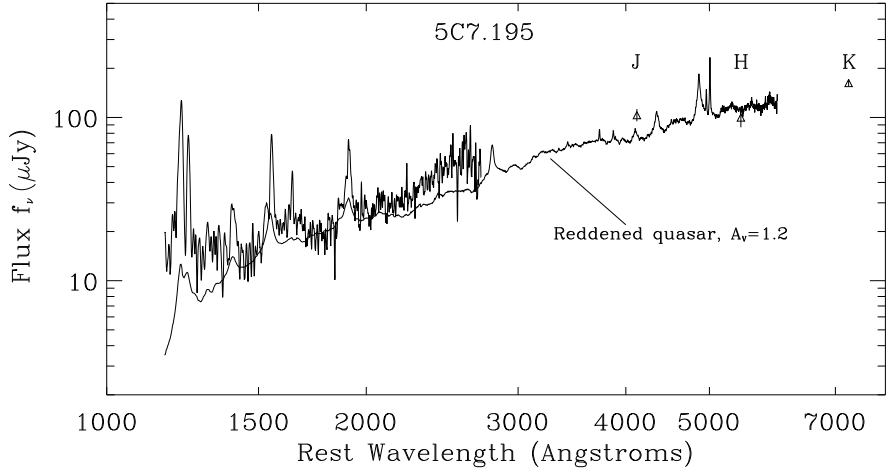,width=12.5cm}
}
\caption{\label{fig} 
Optical spectrum and J, H and K photometry (triangles with error bars) of the red quasar 5C7.195, plotted in the quasars rest-frame. Also plotted here is the LBQS composite quasar which has been artificially reddened ($A{_{V}} = 1.2$), assuming a galactic extinction curve (Savage \& Mathis 1979). } 
\vspace{-0.1cm}
\end{figure}

As seen in section 4, there is very little evidence for reddening amongst the 14 7C quasars with $z < 2$. Three of the seven higher redshift quasars, do appear to be reddened. One of these, 5C7.195 (7C0819+283), has an optical spectrum in which the emission lines are narrow ($FWHM < 2000$ km s$^{-1}$), with only vague hints of weak, broad bases to the C IV $\lambda$1549 and C III] $\lambda$1909 lines. The spectrum is spatially unresolved in both lines and continuum (seeing $FWHM \sim 0.7$ arcsec). The K-band image is also unresolved; PSF subtraction reveals no extended residual down to $K \sim 20.5$. Figure 4 shows the optical spectrum of 5C7.195, J/H/K photometry and a reddened quasar fit. The reddened quasar fits the continuum well except at $\lambda{_{\rm rest}} < 1400$\AA\ and 2200\AA\ $ < \lambda{_{\rm rest}} < 4200$\AA. Note that it is also a good fit to the strengths of the broad emission lines. A weak, flat-spectrum ($\alpha{_{\rm opt}} = 0$) component would account for the discrepancy at $\lambda{_{\rm rest}} < 1400$\AA. The 3000\AA\ bump may be responsible for the poor fit at the long wavelength end of the optical spectrum and the J-band point. We have already seen that this component is much more evident in 7C quasars than in the LBQS composite. An infrared spectrum of 5C7.195 has been obtained in the $2\mu m$ region, which reveals a broad H$\alpha$ line (Willott et al. 1997b). This object illustrates how easy it may be to mis-classify reddened quasars as radio galaxies, unless observations of the objects are taken in good seeing.  

\section{Conclusions}

The 7C Redshift Survey has provided a {\em complete} sample of radio-loud quasars, where the only defining criterion is that $S_{151} > 0.5$ Jy. Slight, positive evolution in the quasar fraction is observed from $0 < z < 1$ to $1 < z < 2$. However, in the redshift range $1 < z < 2$, the quasar fraction appears to be independent of the radio luminosity. These results support a simple orientation model of quasar and radio galaxy unification. The 7C quasars are in general similar to optically-selected quasars. The three objects which are significantly reddened are all at high redshift ($z > 2$).   



\end{document}